# Interface-driven second harmonic generation in indium tin oxide/gold nanolayers: density and mass discontinuities matter

**J. Arya [1] , D. Ramonell [1], J. Trull [1], N. Akozbek [2], M. Scalora [3], C. Cojocaru [1]**

[1]Department of Physics, Universitat Politècnica de Catalunya, Rambla Sant Nebridi 22, 08222 Terrassa (Barcelona), Spain
[2]US Army Space and Missile Defense Command, Technical Center, Redstone Arsenal, Alabama 35898, USA
[3]Institute of Photonics and Quantum Sciences, Heriot-Watt University, Edinburgh, EH14 4AS, UK

## Abstract

Second harmonic generation from transparent conductive oxides and metals is traditionally attributed to Coulomb, magnetic Lorentz, and convective sources. We show that this picture is incomplete in deeply subwavelength indium tin oxide/gold bilayers. Measurements reveal two robust spectral features, a broad maximum near 1000 nm and an epsilon-near-zero peak near 1350 nm, that persist across wide variations in material and excitation parameters. Hydrodynamic-Maxwell simulations identify an interface-specific convective mechanism activated by free-electron density and effective-mass discontinuities at the oxide-metal boundary. As electronic homogenization develops across the bilayer, the response evolves toward one described by conventional Coulomb, Lorentz, and leading convective terms. These results establish second harmonic generation as a sensitive probe of interface-driven nonlinearities and suggest a broader role for electronic discontinuities in shaping nonlinear optical processes at surface and buried material interfaces.

Second harmonic generation (SHG) has long served as a sensitive probe of surfaces and interfaces in centrosymmetric materials. Early studies by Bloembergen and co-workers demonstrated that SHG can occur in reflection from centrosymmetric media when inversion symmetry is broken at a boundary, establishing nonlinear frequency conversion as a powerful diagnostic of interfacial electronic structure [1]. Subsequent work by Shen and others further developed SHG and sum-frequency generation into versatile tools for studying surface and buried-interface properties [2]. In metals and plasmonic structures, hydrodynamic formulations have provided a particularly useful framework for describing SHG through the interplay of Coulomb, magnetic Lorentz, and convective nonlinear sources [3-6]. More recently, transparent conducting oxides (TCOs), especially in their epsilon-near-zero (ENZ) regime, have emerged as a complementary platform for nonlinear optics because of their strong field enhancement and tunable free-electron response [7-9]. Hydrodynamic studies of ultrathin indium tin oxide (ITO) films have shown that Coulomb, Lorentz, convective, nonlocal, and hot-electron effects can produce substantial harmonic generation near the ENZ condition [10]. However, comparatively little attention has been devoted to nonlinear interactions at interfaces formed by adjacent conductive media, where free-electron density and effective mass may both exhibit strong spatial discontinuities [11].

Metal/ENZ oxide heterostructures such as indium tin oxide/gold (ITO/Au) have been studied in the context of nonlinear nanophotonics [7,8], yet their second harmonic response under oblique incidence exhibits features that cannot be fully explained by conventional hydrodynamic models. In Figure 1(a) we schematically show a sample made of 20nm ITO layer deposited on top of gold layers of thickness 100nm and 20nm. Figure 1(b) shows the measured reflected SHG efficiency as a function of the pump wavelength. Both structures display two distinct spectral features: a relative maximum centered near 1000nm and a sharp ENZ-driven absolute maximum near 1350nm, separated by a spectral handover region. For the thicker gold layer, the ENZ peak is roughly an order of magnitude larger than the relative maximum centered near 1000nm maximum, whereas for the 20nm gold layer the discrepancy is reduced to a factor of approximately three.
Standard hydrodynamic descriptions [4-6,10-13], typically formulated for a single conductive medium that employs constant effective mass and no density and mass gradients, attribute SHG to surface Coulomb and magnetic Lorentz forces and predict a monotonic decrease away from the ENZ peak [10]. The relative amplitudes of the SHG efficiency maxima shown in Fig.1(b) cannot be reproduced simultaneously within such constant-mass models. We show that neither conventional Coulomb, Lorentz, nor standard convective sources are sufficient to reproduce the observed ratio between the spectral maxima. Agreement with experiment requires an interface-specific nonlinear source proportional to the combined discontinuities of charge density and effective mass. While density-gradient contributions have been discussed previously [11], the effective-mass discontinuity introduces an additional convective channel that, to our knowledge, has not been identified in the context of SHG from metal/oxide interfaces. The key ingredient is an interface-specific convective term proportional to $\nabla \ln n + \nabla \ln m^*$, activated by the *combined discontinuities of free-electron density and effective mass at the ITO/Au boundary*. The details of the derivation are outlined in the Supplementary Material.

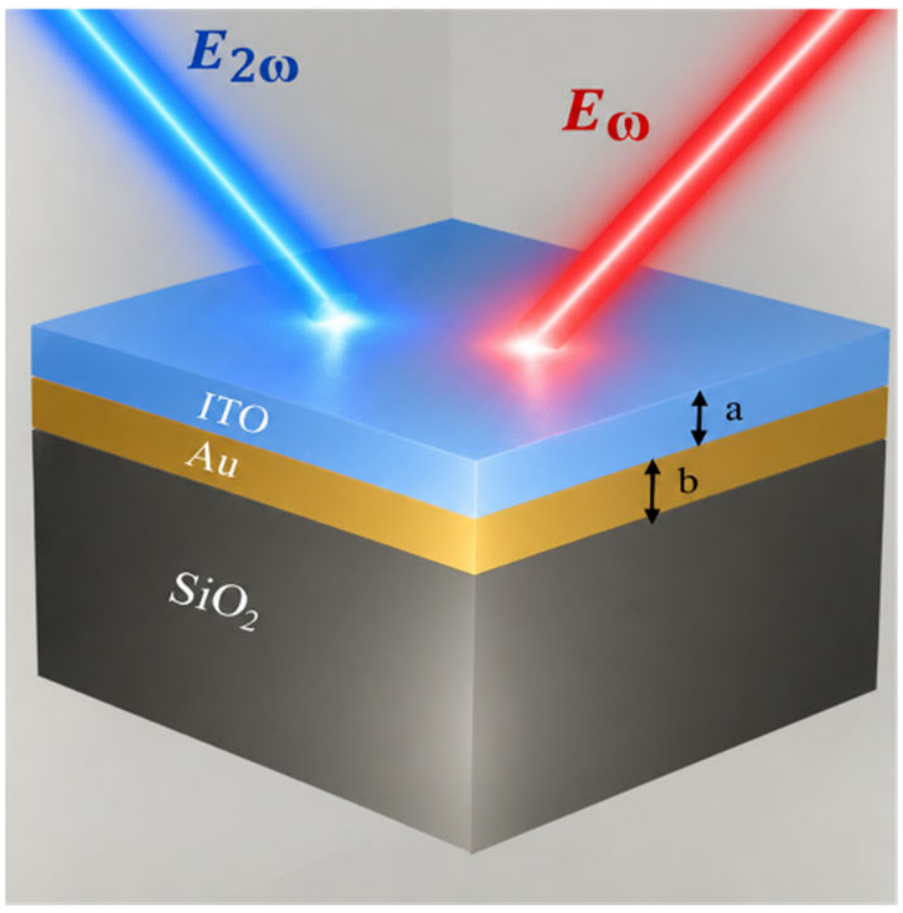

(a)

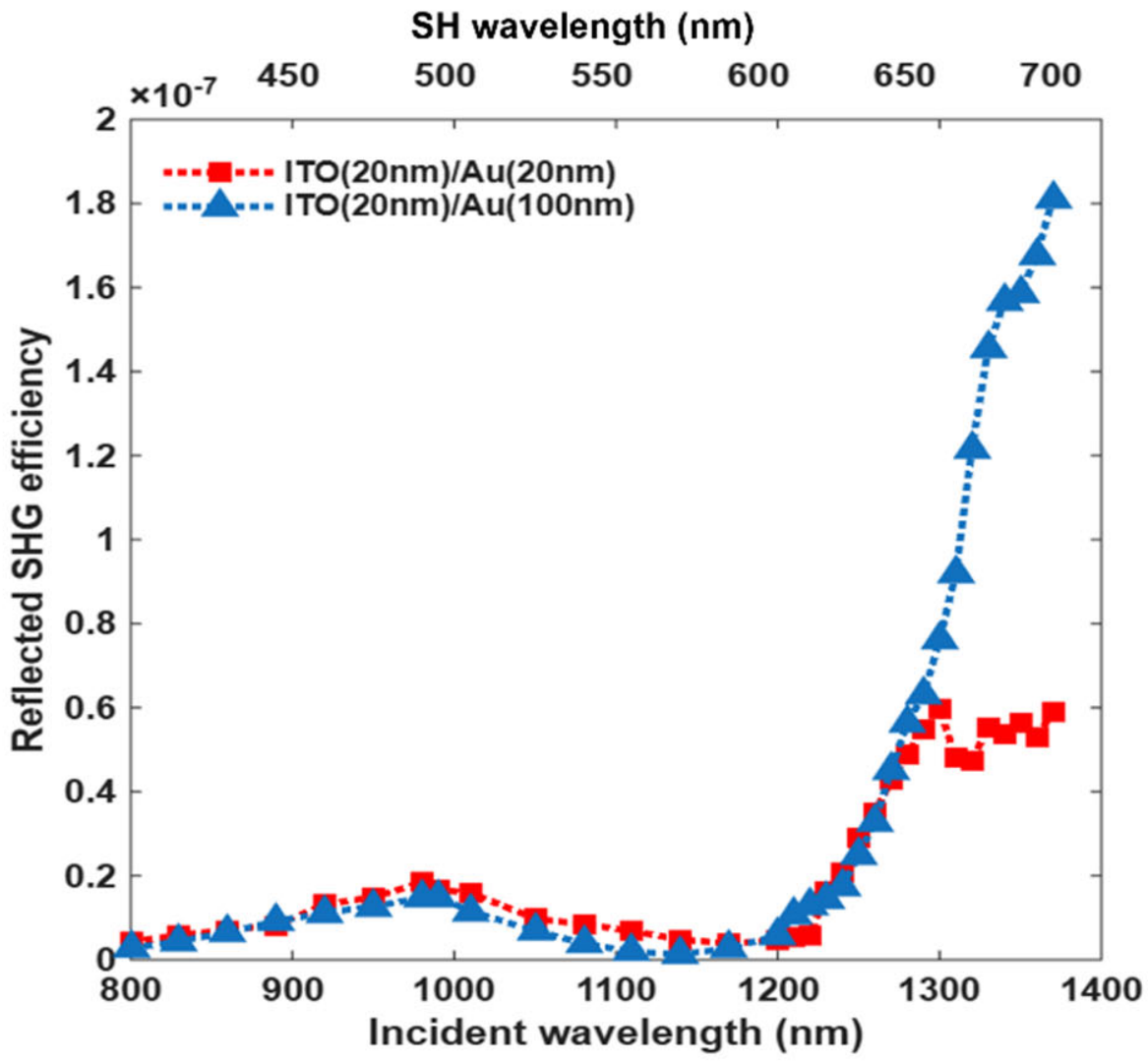

(b)

**Fig.1: (a)** Geometrical representation of a nanometer thick bilayer composed of 20nm thick ITO nanolayer deposited on top of 20nm thick layer of gold. The bilayer rests on top of a glass substrate. **(b)** SHG efficiency measured in reflection from a bilayer composed of a 20nm-thick ITO layer deposited on top of a 100nm-thick (blue curve with full triangular markers) and 20nm-thick (red curve with full square markers) gold layers. Incident peak power density is fixed at approximately 30GW/cm$^2$, with pulse durations ranging from 160fs at the short wavelength range, to 190fs on the opposite side of the spectrum and incident at 60°.

For the *100nm gold layer*, the ITO/Au interface behaves as a *hard electronic wall*: the pump and SH fields do not penetrate the metal, and the discontinuities $\Delta \ln n$ and $\Delta \ln m^*$ contribute at comparable magnitudes. In this regime, the interface convective term is driven by both density and mass jumps, and the bilayer behaves as two distinct electron fluids separated by a sharp boundary. For the *20nm gold layer*, the situation changes dramatically. The metal becomes penetrable, nonlocal effects become important [14,15], the interface behaves more like a diver entering water than a particle striking a wall, and transmitted pump and harmonic fields are substantial. The transverse field profile exhibits cavity-like behavior, the charge distribution smears across each interface, including the exit surface, and both effective-mass and density discontinuities are quenched.

The penetrability of the 20nm Au layer is a well-established property of thin metal films. The dielectric constant of gold at 1350nm is $\varepsilon \approx -77 + i8$ [16], which suggests a huge, highly absorbing potential barrier. However, the classical skin depth describes exponential decay in a semi-infinite conductor; it does not determine transmission through films of finite thickness. The metal behaves as a partially transparent dielectric cavity, with a complex refractive index. This regime underlies the operation of induced transmission filters [17], where single Au or Ag layers as thick as 50nm embedded in dielectric stacks exhibit strong transmission and resonant tunneling. In addition, structures containing hundreds of nanometers of metal, assembled from many layers tens of nanometers in thickness can support high transmission bands due to interference, nonlocality, and boundary induced suppression of absorption [18]. Thus, the concept of skin depth loses its meaning in layered systems, where the effective absorption coefficient is dramatically reduced and the metal becomes optically penetrable. Our 20nm Au layer lies in this regime. As a result, the ITO(20nm)/Au(20nm) bilayer behaves as a single, deeply subwavelength electron fluid with increased elasticity and reduced density and mass discontinuities. In this deeply subwavelength 40nm bilayer, the hydrodynamic-Maxwell model indicates that a single effective mass can be used throughout the structure. The nominal density discontinuity remains, but it is smoothed by an active charge density redistribution across the volume.

This transition from a sharply discontinuous interface to a spatially homogenized plasma can be understood through an analogy with quantum heterostructures. In microscopic quantum mechanics, an abrupt spatial step-change in effective mass across a boundary requires the implementation of BenDaniel-Duke boundary conditions [19]. These conditions modify the spatial derivative of the wavefunction ($\frac{1}{m^*}\nabla\Psi$) to enforce the continuity of the probability flux across the mass mismatch. Interestingly, the core mathematical structure that ensures this quantum probability current is conserved under localized mass gradients mirrors the spatial gradient term in the pulse-envelope Poynting vector **S** that governs local energy flow [20]. In both frameworks, the transport of a conserved quantity, whether quantum probability or classical electromagnetic energy, relies on an identical mathematical construct that couples a complex field envelope with the spatial gradient of its complex conjugate. This term tracks the anomalous, phase-modulated current generated when waves encounter a strong barrier. On a macroscopic scale, our hydrodynamic framework (Eq. S7, supplementary material) acts as the fluid-dynamics counterpart to this quantum constraint, where the explicit $\nabla \ln m^*$ term forces the velocity fields to adjust to the localized inertia jump at a "hard" electronic boundary. When the gold layer is thinned to 20nm, nonlocal field penetration and charge smearing effectively blur this boundary. In quantum mechanics, this is akin to transitioning from an abrupt step-potential to a smoothly graded heterojunction, where the BenDaniel-Duke interface constraints naturally vanish as the gradient softens. Consequently, the quenching of the $\nabla \ln n$ and $\nabla \ln m^*$ terms in the 20nm/20nm bilayer reflects a macroscopic "smoothing" of the electronic flux constraints, allowing the structure to shed its boundary-dominated character as it smoothly transitions to a homogenized effective medium.

A more physical interpretation is that intense pump excitation activates deeper, more strongly bound electronic states at SH wavelengths, in spectral regions where the Lorentzian response begins to dominate (below ~500 nm). These valence-derived carriers contribute to the effective free-electron density but possess heavier band masses and shorter coherence times. In hydrodynamic language, this corresponds to wavelength- and intensity-dependent effective-mass renormalization: absorption, carrier heating, and local-field enhancement conspire to modify the inertia of the electron fluid, producing additional nonlinear damping terms proportional to $(1/n)\partial_t n$ and $(1/m)\partial_t m$ [21], thus imposing a kind of temporal inertial brake on the system. Simulations suggests that in the thicker bilayer, near the ENZ crossing point, the local pump field inside ITO is enhanced by a factor of four, causing significant effective-mass excursions that attenuate the nonlinear channels that scale as $1/m^*$. Thus, the conventional Coulomb, Lorentz, and convective contributions remain active at long and short wavelengths, but their reduced strength allows the interface convective term to become comparatively more influential. The mathematical derivation that leads to the additional convective terms associated with gradients of the equilibrium electron density and effective mass is given in the Supplementary Material. The discussion below focuses on the implications of these terms for the nonlinear optical response and, in particular, on the extent to which they are required to reproduce the measured SHG spectra.

In the ITO(20nm)/Au(100nm) system considered here, both $\Delta \ln n_0$ and $\Delta \ln m^*$ are significant, and their sum controls the strength and sign of the interface nonlinear response. For typical ITO → Au direction of propagation, with respective densities of order $2 \times 10^{21}$ cm$^{-3}$ and $5.8 \times 10^{22}$ cm$^{-3}$, respectively, the fitting procedure yields $\Delta \ln n \approx 4$ and $\Delta \ln m^* \approx -2.8$. These magnitudes should be regarded as effective parameters that account for interface transport and nonlocal charge redistribution at the SH wavelength. While not directly measured, they remain physically plausible and consistent with the observed nonlinear response and correspond to assigning a heavier effective mass to ITO (~1.5 $m_e$) and a lighter one to Au (~0.085 $m_e$) at the second harmonic wavelength. This is physically plausible: interface roughness and scattering can increase the inertia of longitudinal currents in ITO, while nonlocal "softening" and subsequent smearing of the charge across the Au immediate sub-surface, can reduce scattering and favor lighter-mass transport pathways.

The situation is dramatically different in the ITO(20 nm)/Au(20 nm) case. The respective densities of Au and ITO are chosen to be $5.8 \times 10^{22}$ cm$^{-3}$and $1.25 \times 10^{22}$ cm$^{-3}$, with effective masses $m^*_{ITO,Au} \approx 0.125\ m_e$. When the metal layer is thin and penetrable, nonlocal effects, which are fundamentally strong near surfaces, are easily triggered in the gold section. The entire 40 nm-thick stack behaves as a deeply subwavelength electron fluid. The hard-wall character of the thick metal is replaced by a "soft interface" where charge density smears across the boundary, effective-mass contrast is reduced, and the spatial derivatives of fields and polarization are correspondingly weakened. This leaves the first two convective terms in Eq. S9 (supplementary material) to contribute most. The net result is a reduced interface convective source and lower SHG efficiency, as can be ascertained by comparing measured and simulated reflected SHG amplitudes in Fig.2. The figure shows that both the relative and the absolute maxima are reproduced simultaneously, provided that in each case the phenomenology is described by appropriate effective masses and densities.

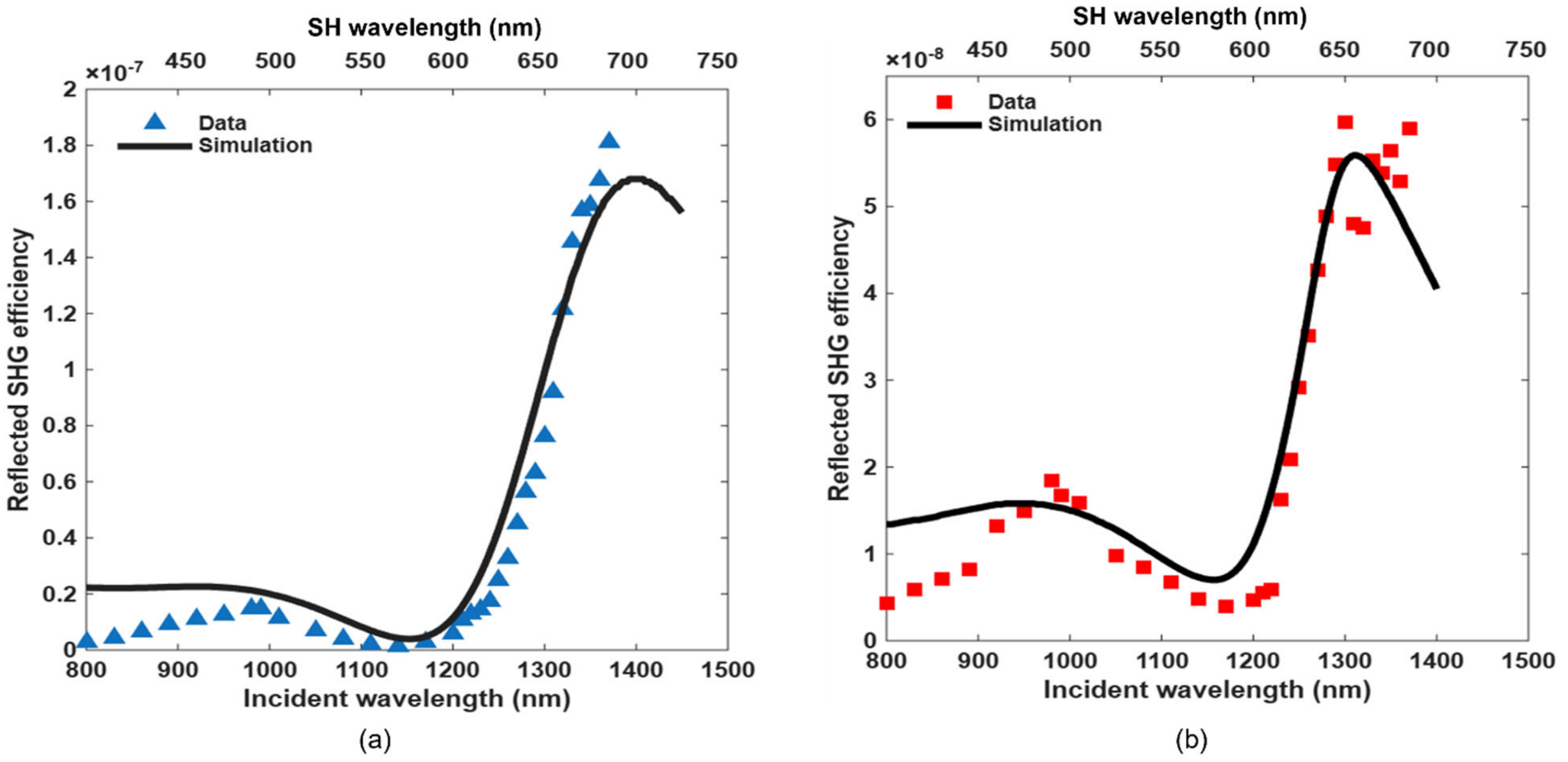

**Fig.2:** Measured (solid blue triangles and red squares) and simulated (solid curve) reflected SHG efficiency from a bilayer composed of a 20nm-thick ITO layer deposited on top of: **(a)** 100nm-thick and **(b)** 20nm-thick gold layers. The hydrodynamic-Maxwell model that we use captures both qualitative and quantitative aspects of the dynamics quite well, but some discrepancy remains at short wavelengths, where simulations predict slightly higher efficiencies than measured. Although not shown here, agreement can be restored by allowing the effective masses to renormalize, increasing for Au and decreasing for ITO, as the pump wavelength is tuned away from the ENZ crossing point.

To extract the physical origin of the two peaks in the SHG spectra, we perform two different simulations. In the first we will consider the individual effects of metal and TCO on SHG. In Fig.3 we display simulated reflected SHG efficiency from a gold mirror 100nm thick, and from a 20nm-thick ITO layer. It is evident that the relative maximum centered near 1000nm seen in Fig.2 originates in the metal nonlinearity. It is equally apparent that the SH peak that develops near 1350nm is triggered by the ITO. Coulomb, magnetic Lorentz and convective terms are all active in both materials. However, for individual isolated layers the convective sources in Eq.S9 (supplementary material) cannot contain either density or effective mass gradient terms. The results shown in Fig.3 confirm what has already been reported separately in reference [13] for single gold layers (see especially Fig.2 in ref.[13]), and in reference [10] for a single 20nm-thick ITO layer.

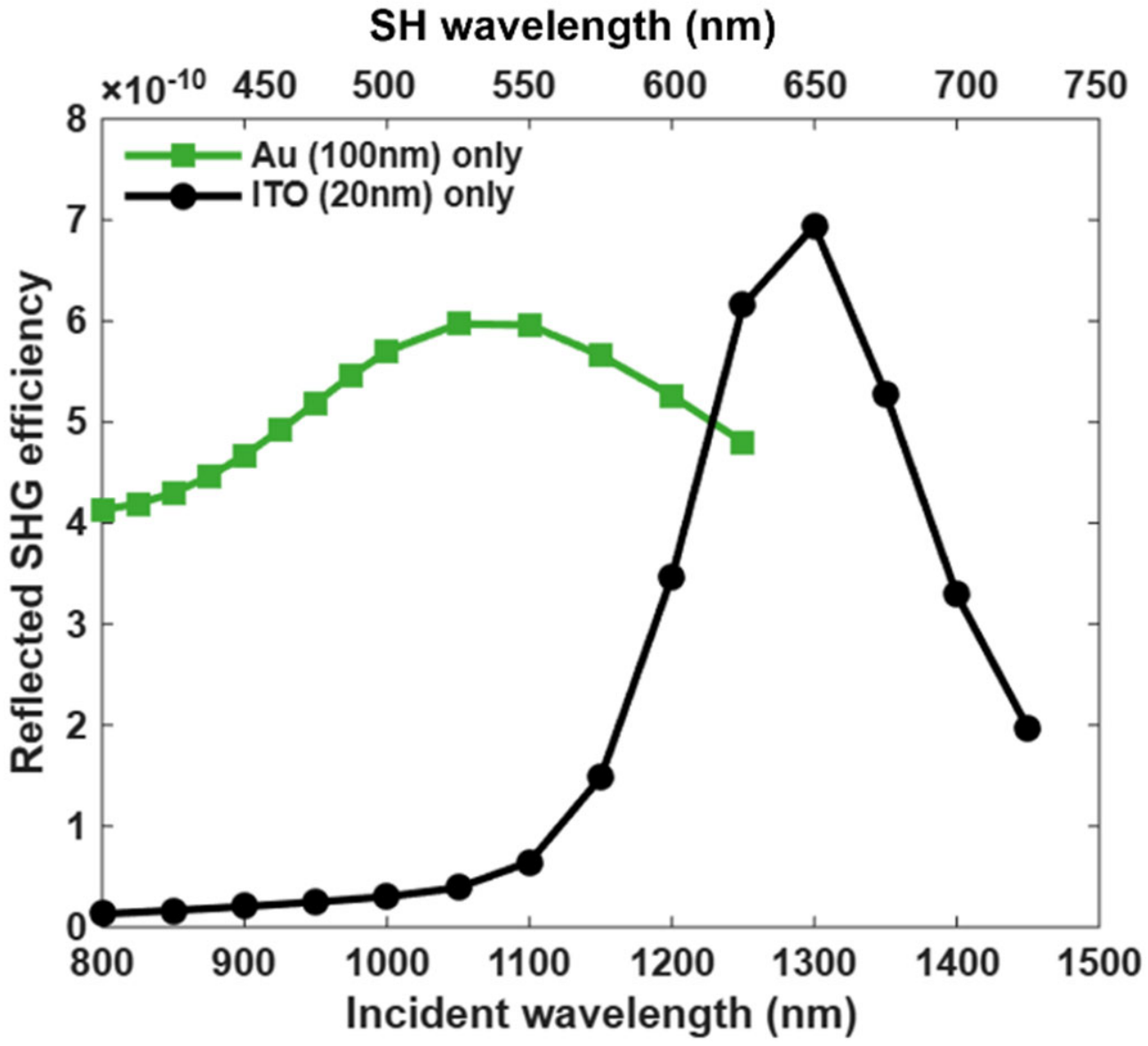

**Fig.3:** Reflected SHG efficiencies from isolated 20nm-thick ITO and 100nm-thick gold layers, as predicted by the hydrodynamic-Maxwell simulations, using constant effective masses and without charge density or mass gradients or discontinuities.

In the second simulation we recalculate the SHG efficiency spectrum for the ITO(20nm)/Au(100nm) (Fig.2a) by arbitrarily setting the density and mass gradient terms shown on the right hand side of Eq.S9 (supplementary material) equal to zero: $\nabla \ln n_0 = \nabla \ln m^* = 0$. Figure 4 constitutes perhaps the clearest evidence for the interface mechanism proposed here. When the density- and mass-gradient contributions are artificially suppressed, the model immediately loses the spectral handover region and predicts a nearly monotonic response away from the ENZ wavelength. The experimentally observed coexistence of the metal-related maximum and the ENZ peak therefore appears to require the interface convective source. Hence the importance of assessing and including density and mass discontinuities in the description of the interaction. The interface mechanism that we report thus explains the spectral handover observed in Figs. 1 and 2: at short wavelengths, SHG is dominated by gold-based surface and Lorentz forces, whereas at longer wavelengths, effective-mass renormalization in ITO favors interface convective terms that produce a handover region and a large ENZ peak near 1350 nm.

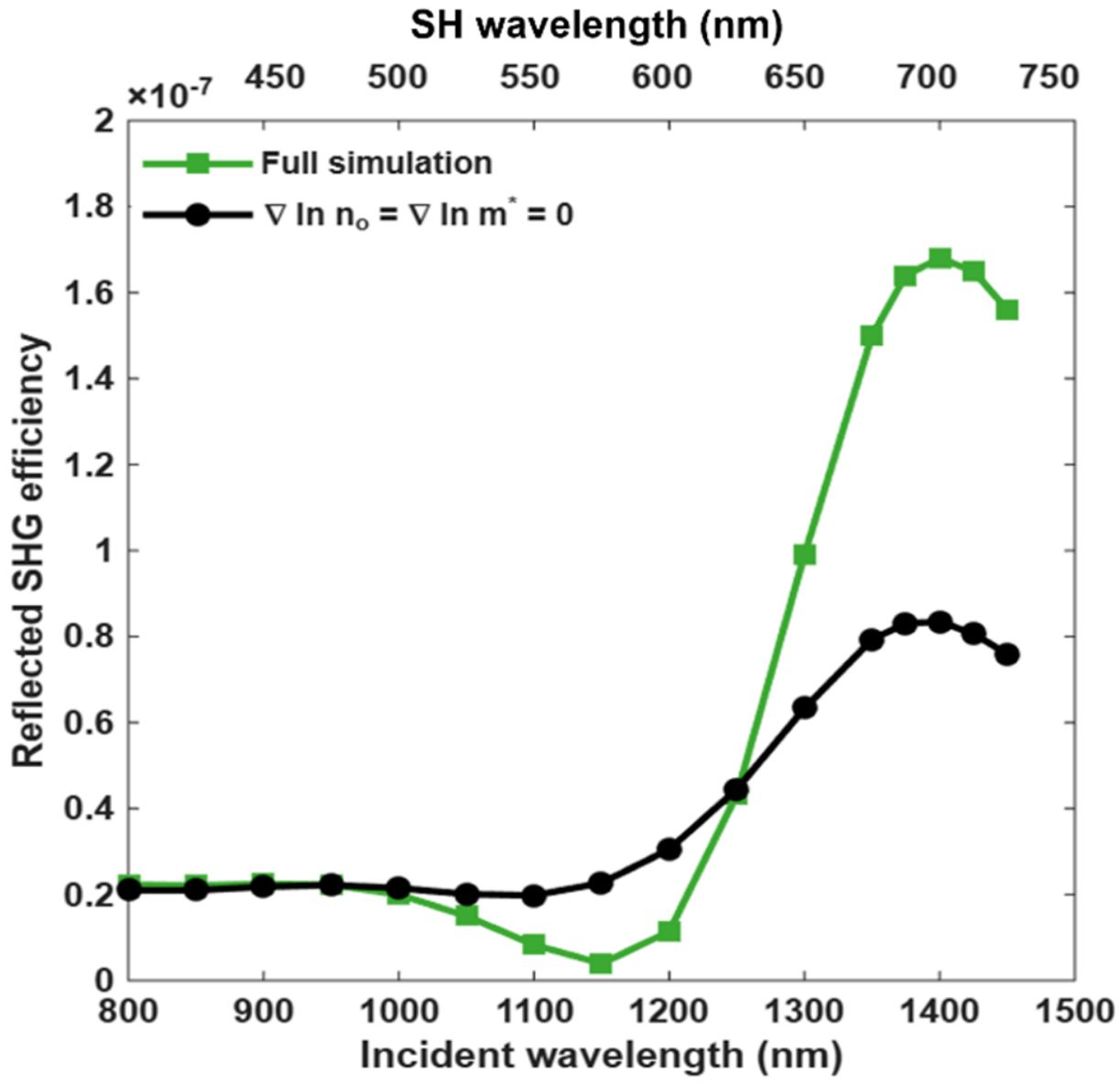

**Fig.4:** Reflected SHG efficiencies from the ITO(20nm)/Au(100nm) bilayer reported in Fig. 2a (full simulation), as indicated in the caption. Excluding the gradient terms in the convective portion of the nonlinear sources leads to monotonically decreasing SHG efficiency away from the ENZ condition and fails to yield the observed ratio between relative SHG efficiency maxima.

To further test the physical picture developed above, we investigated the reversed bilayer geometry consisting of a 20nm-thick Au layer deposited on a 20nm-thick ITO film. Unlike the ITO(20nm)/Au(100nm) system, where a sharp interface convective source is required to reproduce the measured spectrum, the reversed Au(20nm)/ITO(20nm) structure exhibits behavior that is more closely related to the thin ITO(20nm)/Au(20nm) bilayer discussed previously. Experimental measurements and hydrodynamic-Maxwell simulations displayed in Fig.5 show good agreement across the entire spectral range, including the relative amplitudes and positions of the observed SHG features. To assess reproducibility, the SHG spectrum was measured independently on different days under nominally identical conditions. The datasets, shown using different symbols, exhibit moderate run-to-run variability while retaining similar overall spectral dependence, general shape and order of magnitude. The theoretical curve passes through the middle of the experimental distribution, reproducing the principal spectral features observed in all measurements. The most notable aspect is that the simulations were carried out using identical effective masses in both materials, and without including explicit density- and effective-mass-gradient terms that appear in the convective nonlinear source (Eq. S9, supplementary material). The calculation therefore does not rely on interface- specific contributions proportional $\nabla \ln n_0$ and $\nabla \ln m^*$. We note that actual layer thicknesses may be slightly different from the nominal 20nm we have used for both ITO and Au, and there may also be differences in dielectric constants between the Drude-Lorentz fit and Palik's data. Notwithstanding that, the measured SHG spectrum is reproduced with reasonably good accuracy.

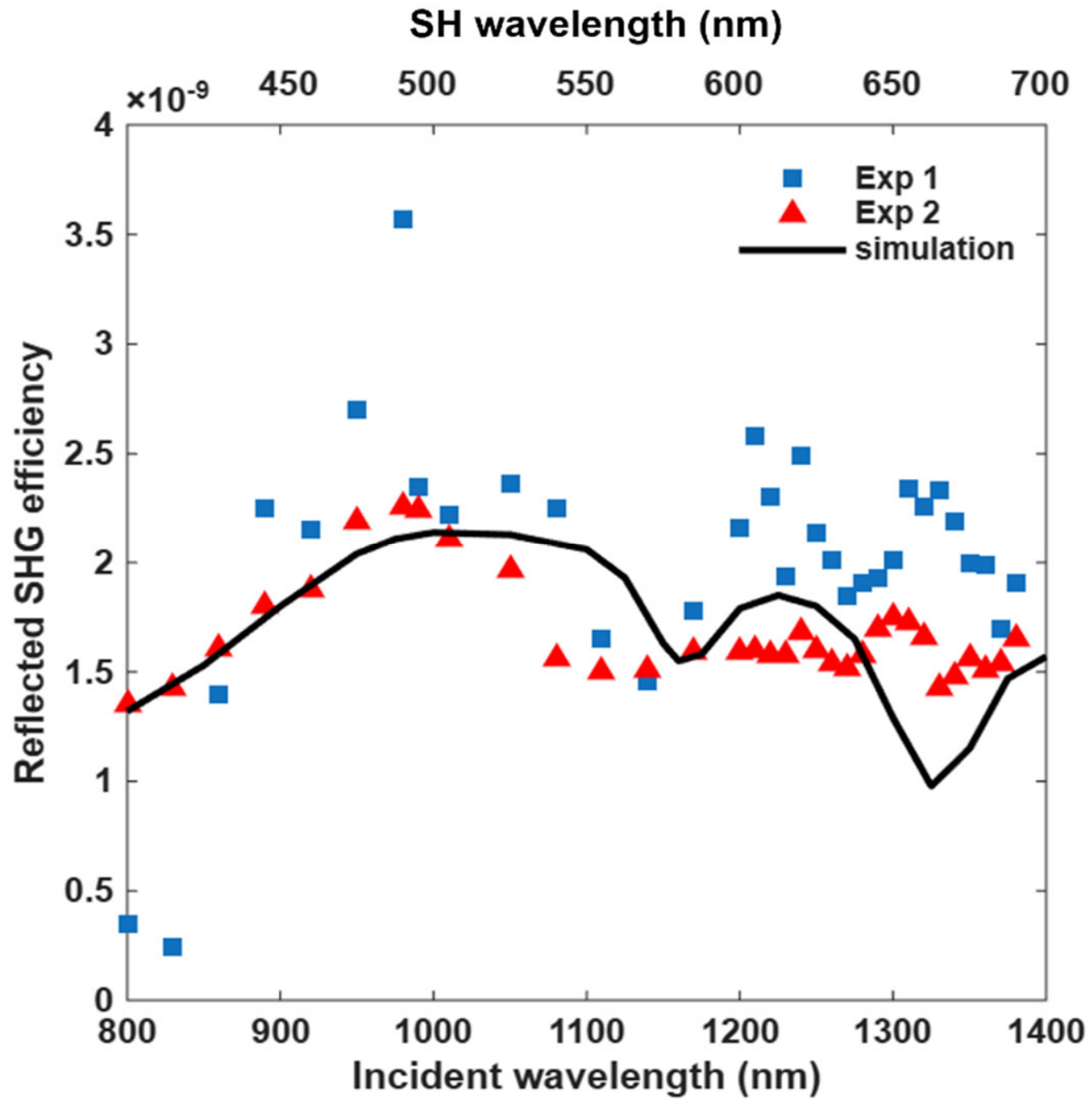

**Fig. 5:** Measured (solid and blue squares and red triangles) and simulated (solid black curve) reflected SHG efficiency from an Au(20nm)/ITO(20nm) bilayer. Simulations use $m^*_{Au,ITO} \approx 0.28m_e$ and exclude explicit density- and mass-discontinuity contributions in the convective sources.

This agreement should not be interpreted as evidence that the free-electron density is continuous across the Au/ITO interface. A substantial density mismatch remains present. Rather, the result indicates that the nonlinear dynamics are dominated by the conventional hydrodynamic contributions, namely the Coulomb, magnetic Lorentz, and the leading convective terms involving current-density gradients. The explicit discontinuity-driven terms appear to play only a minor role for this structure. In other words, the density discontinuity exists physically but does not manifest itself as a dominant nonlinear source. This observation is consistent with the effective-medium picture introduced earlier for deeply subwavelength conductive bilayers. Because both layers are only 20nm thick, pump and harmonic fields penetrate the entire stack and sample both materials simultaneously. Nonlocal charge redistribution and finite transmission through the thin Au layer reduce the sensitivity of the nonlinear response to the abrupt microscopic interface, even though the underlying carrier densities remain different. From the perspective of the hydrodynamic equations, the bilayer behaves approximately as a single electron fluid characterized by a nearly uniform effective mass.

The reversed Au/ITO geometry therefore provides an independent validation of the broader interpretation developed in this work. The hydrodynamic-Maxwell framework remains fully capable of describing the measured SHG response, but the dominant nonlinear mechanism depends on the effective electronic character of the interface. In thick, electronically abrupt structures, discontinuity-driven convective sources are essential for quantitative agreement. In thin, optically penetrable bilayers, the first conventional convective terms become dominant and the response approaches that of a homogenized plasma with renormalized parameters. The good agreement obtained for the Au(20nm)/ITO(20nm) bilayer and its reversed counterpart therefore reinforces the conclusion that the relevant question is not whether a density discontinuity exists, but whether that discontinuity acts as a dominant nonlinear source for the optical process under consideration.

The behavior of the ITO/Au bilayers studied here, particularly the transition from a hard, sharply discontinuous interface in the 20 nm/100 nm structure to a soft, penetrable interface in the 20 nm/20 nm structure, has conceptual parallels with recent developments in time-varying [22] ENZ films [21,23]. In both cases, the underlying physics is governed not by the detailed microscopic structure of the material, but by the emergence of an effective plasma whose parameters evolve either in space or in time. This connection provides a broader framework for understanding when multilayers of transparent conducting oxides and thin metals behave as distinct media and when they merge into a single, homogenized electron fluid.

In the ITO(20nm)/Au(100nm) case, the pump and SH fields penetrate only a few nanometers into the gold, and the discontinuities in free-electron density and effective mass remain sharp. The interface convective term is therefore strongly localized and driven by both $\Delta \ln n_0$ and $\Delta \ln m^*$, producing the large ENZ-enhanced SHG peak near 1350nm. This regime is analogous to the static, spatially abrupt boundary considered in conventional hydrodynamic models, where the nonlinear response is dominated by surface Coulomb, Lorentz, and convective terms that arise from sharp spatial gradients.

In contrast, the ITO(20nm)/Au(20nm) bilayer behaves very differently. The metal layer is thin enough for nonlocal effects to become prominent, and the fields penetrate the entire 20nm gold slab. The charge density smears across the boundary, the effective-mass discontinuity is softened, and the spatial derivatives of the fields and polarizations are reduced. In this regime, the entire 40 nm stack behaves as a single deeply subwavelength electron fluid, and the hydrodynamic-Maxwell model suggests that a uniform effective mass can be used throughout the bilayer. The interface convective term is still present, but it is now dominated primarily by spatial derivatives of the current densities as its overall strength is reduced, consistent with the lower SHG efficiency observed in Fig. 1(b).

This spatial homogenization has a direct conceptual analogue in the time-domain homogenization observed in recent time-varying ENZ films. In Ref. [21], AZO is pumped so that its plasma parameters, effective mass, damping, and carrier distribution, evolve rapidly in time. The film behaves not as a stack of interfaces but as a single time-varying plasma, whose anisotropy and optical activity arise from ultrafast symmetry breaking in the electron fluid. Ref. [23] similarly treats a 900 nm AZO film as a uniform, non-stationary medium whose refractive index changes in time, producing spectral shifts, angular deflection, and pulse fission through large temporal gradients $dn/dt$. In both works, the material is modeled as a homogeneous electron gas with renormalized, time-dependent parameters, rather than as a multilayer with spatially distinct boundaries.

Our thin-metal bilayer exhibits a spatial counterpart of this behavior. When the gold layer is sufficiently thin and transparent, according to measurements that are supported by simulations, the bilayer ceases to act as two separate media and instead forms a spatially homogenized effective plasma [24]. The physical boundaries blur, the electron density and effective mass become continuous across the interface, and the nonlinear response is governed by renormalized hydrodynamic parameters, precisely the same conceptual structure used to describe time-varying ENZ films. This suggests that adjacent low-index or ENZ materials such as ITO/AZO, ITO/CdO,

TiN/ITO, and even a bilayer formed by two identical TCOs annealed at different temperatures and displaying different dielectric response, when fabricated at thicknesses below ~20-40 nm, may behave as a *single effective medium* with unified plasma parameters and reduced interface contributions. In such systems, the hydrodynamic-Maxwell model naturally transitions from a boundary-dominated description to an effective-medium description where discontinuities still matter, but the dynamics mirrors the transition from static to time-varying plasmas in the AZO studies.

Thus, the ITO/Au bilayers studied here occupy a conceptual connection between interface-driven nonlinear optics and time-varying plasma optics. Thick metals preserve sharp boundaries and strong interface convective sources, while thin metals enable spatial homogenization and effective-medium behavior. This duality provides a unique perspective on nonlinear ENZ photonics, linking spatially structured bilayers with temporally structured ENZ films and highlighting the broader principle that nonlocality, transparency, and subwavelength thickness can merge multilayer plasmas into a single, renormalized electron fluid.

The reversed Au(20nm)/ITO(20nm) bilayer provides additional support for this interpretation. While the ENZ resonance on the ITO side is shielded, and despite the presence of a substantial free-electron density discontinuity at the metal-oxide boundary, the measured SHG spectrum is reproduced well using nearly identical effective masses in both materials and without including the explicit density- and mass-gradient terms in the convective sources. This result demonstrates that the existence of a density-discontinuity does not necessarily imply that discontinuity-driven nonlinear sources dominate the optical response. Rather, for sufficiently thin and optically penetrable bilayers, the first conventional convective contributions together with the Coulomb and Lorentz terms appear to govern the dynamics, while the explicit interface-gradient terms become secondary. The good agreement obtained for the reversed geometry therefore reinforces the notion that deeply subwavelength metal/oxide stacks can transition toward an effective-medium description characterized by a nearly uniform electron-fluid response. In addition, this agreement should not be interpreted literally as evidence that the microscopic effective masses of Au and ITO become identical. Rather, the result points toward a more subtle picture in which the nonlinear optical response becomes insensitive to fine-scale electronic inhomogeneities that remain physically present at the interface. In fact, there is evidence that interfacial electronic properties are inherently variable. Then Fig. 5 becomes the modern nonlinear-optics manifestation of a problem that the thin-film and plasmonic communities have been observing for decades.

This distinction is important because the electronic properties of thin metallic films are known to differ substantially from their ideal bulk values, and to depend sensitively on thickness, growth conditions, and local environment. For example, Dryzek et al. [25] reported significant variations of the extracted free-electron parameters of sputtered noble-metal films as a function of both film thickness and deposition technique, with measurable changes in effective mass, scattering length, and other transport parameters. Likewise, Nash and Sambles [26] demonstrated that the optical response of a single silver film may differ markedly at its two opposite interfaces, with the metal/vacuum and metal/substrate boundaries exhibiting distinct dielectric functions, plasma frequencies, and effective free-electron characteristics. Their analysis further suggested that porosity, density gradients, surface contamination, and local growth morphology can all contribute to the formation of electronically distinct interfacial regions.

Viewed in this context, the conventional picture of an abrupt junction separating two perfectly homogeneous bulk materials may represent an oversimplification, particularly for films whose thickness is comparable to optical penetration depths and nonlocal length scales. The Au/ITO and ITO/Au interfaces likely contain a spatially varying electronic environment characterized by gradual changes in carrier density, damping, and effective mass that develop naturally during growth and processing. Such variations are entirely consistent with the physical quantities that appear explicitly in the hydrodynamic equations through gradients of the equilibrium density and effective mass.

The significance of Fig. 5 is therefore not that electronic inhomogeneity has disappeared. On the contrary, the historical thin-film literature suggests that some degree of inhomogeneity should be expected. Indeed, the measurements indicate that the nonlinear optical process no longer resolves those microscopic details strongly enough for the corresponding discontinuity-driven terms to dominate the response. The electron-density discontinuity remains present, and local variations of plasma frequency, damping rate, and effective mass may persist across the interface. Nevertheless, the measured SHG spectrum is described remarkably well by a model that behaves as though the two layers participate in a common electronic response.

This observation supports the view that the Au(20nm)/ITO(20nm) structure occupies a different physical regime from the ITO(20nm)/Au(100nm) bilayer. In the latter case, the nonlinear response retains a strong sensitivity to interface-specific electronic structure, and the density- and mass-gradient contributions emerge as dominant sources. In the former case, the same discontinuities remain physically present but become increasingly embedded within a distributed electronic environment that is sampled collectively by the optical field. The resulting dynamics are therefore more appropriately described in terms of a nearly homogenized electron fluid characterized by effective parameters rather than by sharply separated material domains.

These findings suggest that the relevant question is not whether density and effective-mass discontinuities exist at a given interface, but rather whether the optical process under consideration is capable of resolving them. The transition observed between the thick and thin bilayer geometries may therefore represent a transition from an interface-resolved regime to an effectively homogenized regime, even though the microscopic electronic structure remains nonuniform throughout.

In conclusion, our results show that SHG in ITO/Au and Au/ITO bilayers is governed by a hydrodynamic interplay between conventional Coulomb, Lorentz, and convective sources and, when present, interface-specific convective mechanisms arising from discontinuities in free-electron density and effective mass at the metal-oxide boundary. Thick, electronically abrupt interfaces require explicit discontinuity-driven contributions, whereas thin and optically penetrable bilayers can be described accurately using a nearly homogenized electron-fluid picture with renormalized parameters. Conventional Coulomb, Lorentz, and convective sources remain active across the spectrum, but modest wavelength- and intensity-dependent effective-mass renormalization, driven by absorption, carrier heating, and local-field enhancement, attenuates their contributions just enough for the interface term to dominate the nonlinear response. These combined mechanisms explain the spectral features of the SHG signal, which include a peak near 1000nm, a handover region centered near 1150nm, the large ENZ peak near 1350nm, and resolves discrepancies between experiment and constant-mass hydrodynamic models. These results show that while the hydrodynamic-Maxwell framework provides the correct structural form for nonlinear sources, the frequency-resolved masses and damping rates that enter the model must be treated as renormalized, phenomenological quantities, highlighting interface-driven nonlinearities as a key ingredient in conductive-oxide/metal nanophotonics. The results thus suggest that SHG can serve as a sensitive probe of electronic discontinuities at buried interfaces [27] and indicate that nonlinear optical measurements may provide direct access not only to surface charge dynamics, but also to effective-mass variations that are otherwise difficult to characterize experimentally.

**Acknowledgements**

J.A., J.T., and C.C. acknowledge the Spanish Agencia Estatal de Investigación from Ministerio de Ciencia, Investigación y Universidades (project PID2023-148620NB-I00 granted by MICIU/AEI/10.13039/501100011033 and FEDER, UE). The authors thank M.A. Vincenti, D. de Ceglia, M. Ferrera, W. Jaffray, S. Stengel, and C. Rizza for critical reading of the manuscript.